\documentclass[conference]{IEEEtran}
\IEEEoverridecommandlockouts
\usepackage{cite}
\usepackage{hyperref}
\usepackage{amsmath,amssymb,amsfonts}
\usepackage{algorithmic}
\usepackage{graphicx}
\usepackage{textcomp}
\usepackage{xcolor}
\def\BibTeX{{\rm B\kern-.05em{\sc i\kern-.025em b}\kern-.08em
    T\kern-.1667em\lower.7ex\hbox{E}\kern-.125emX}}
\begin{document}

\title{Feasibility on Detecting Door Slamming towards Monitoring Early Signs of Domestic Violence}

\author{\IEEEauthorblockN{Osian Morgan}
\IEEEauthorblockA{\textit{School of Computer Science and } \\
\textit{Informatics, Cardiff University}\\
Cardiff, UK \\
MorganOB@cardiff.ac.uk}
\and
\IEEEauthorblockN{Hakan Kayan}
\IEEEauthorblockA{\textit{School of Computer Science and } \\
\textit{Informatics, Cardiff University}\\
Cardiff, UK \\
KayanH@cardiff.ac.uk}
\and
\IEEEauthorblockN{Charith Perera}
\IEEEauthorblockA{\textit{School of Computer Science and } \\
\textit{Informatics, Cardiff University}\\
Cardiff, UK \\
PereraC@cardiff.ac.uk}
}

\maketitle

\begin{abstract}
  By using low-cost microcontrollers and TinyML, we investigate the feasibility of detecting potential early warning signs of domestic violence and other anti-social behaviors within the home. We created a machine learning model to determine if a door was closed aggressively by analyzing audio data and feeding this into a convolutional neural network to classify the sample. Under test conditions, with no background noise, an accuracy of 88.89\% was achieved, declining to 87.50\% when assorted background noises were mixed in at a relative volume of 0.5 times that of the sample. The model is then deployed on an Arduino Nano BLE 33 Sense attached to the door, and only begins sampling once an acceleration greater than a predefined threshold acceleration is detected. The predictions made by the model can then be sent via BLE to another device, such as a smartphone of Raspberry Pi.
\end{abstract}


\section{INTRODUCTION AND MOTIVATION}

With the ever-increasing availability of low-cost microcontrollers and other computing devices, and advances in more lightweight machine learning techniques, it is becoming increasingly viable to make many of the everyday objects found in our homes smarter. In this study we looked at the feasibility of using inexpensive microcontrollers to leverage machine learning techniques to detect specific phenomena. In this case, we look at the possibility of using TinyML and an Arduino Nano BLE 33 Sense to detect whether a door is being slammed shut aggressively or being shut normally. 

In a similar study\cite{b3}, we see how the same Arduino board, in conjunction with the Edge Impulse platform can be used to detect the wing beats of Mosquitoes reasonably accurately.

The COVID-19 pandemic resulted in many changes and restrictions for our daily lives, most notable being mandates to work from home where possible, as well as legal requirements to socially isolate. Between March 2020 and June 2020, police in England and Wales recorded a 7\% increase in offences flagged as domestic abuse related, with the ONS noting a general increase in demand for domestic abuse victim support services (including a 65\% increase in calls to the National Domestic Abuse Helpline between April and June 2020, compared to the previous quarter). Overall, the entire 12-month period between March 2020 and March 2021 saw an overall increase of 6\% in domestic abuse related crimes.\cite{b5} This follows general increases seen in previous years and could be associated with increased reporting by victims, in addition to improved recording by police forces. \cite{b1}

With this in mind, we seek to demonstrate the viability of using microcontrollers embedded within doors in the home, to assist social workers and law enforcement in the monitoring of potentially aggressive behaviors and perhaps the early signs of domestic violence in social housing.

\section{Our Approach and Methods }
We began by breaking the problem down into its component parts. Namely what physical properties can we measure to determine if a door has been slammed, how do we decide if a sample shows a slam or a normal close, and how could we implement this in a way that’s practical for local authorities, social workers and law enforcement?

\subsection{How to quantify a closing door?}

We decided that accelerometer and audio data were the most useful for this task. When thinking about how a human would decide if a door is being closed aggressively or not, there are typically two physical parameters that spring to mind; how fast the door is closing, and how much sound is generated when it hits the frame. We then selected a suitable microcontroller for the task, we opted for the Arduino Nano BLE 33 Sense, as this small board ships with the required instruments as standard, in addition to having BLE capability. Next, we experimentally determined that a 2 second sampling window was sufficient for slamming standardized MDF interior doors.

By recording variations in acceleration in three spatial dimensions (x, y, z) we can see notable differences between a slam and a normal close. We recorded this data using the IMU on-board the Arduino and sent the samples to Edge Impulse, a cloud-based web-platform for building and deploying machine learning models for resource constrained devices, such as our Arduino.

After collecting accelerometer data we moved on to collecting audio samples, each sample once again being 2 seconds in length. As shown in a 2021 study \cite{b4}, we can see that Arduinos are capable of handling audio analysis without the need for a PC, thus keeping computing as close to the edge as possible and allowing for their devices to be kept portable. In our case, this would mean we'd be better able to integrate our devices into the build environment without the need for larger more conspicuous computers. The audio signal was less straightforward to differentiate, however when visualized using librosa (a python library for handling audio data), we could visually see marked differences in the spectrograms produced by slams as compared to those of normal closes. Much like the accelerometer samples, these were then uploaded to Edge Impulse to begin work on a model. 

\subsection{Building a classification model}

Once we uploaded the data, we began to look at possible features we could use to train a neural network. For accelerometer this is would be relatively straightforward as the many X, Y, Z values of each axis generated per sample can be flattened down into simple 3D coordinate. In this case we took the root mean squared of X, Y, Z and plotted these on a 3-dimensional grid. The result being easily identifiable clusters of slam and normal samples. 

\begin{figure}
    \centering
	\includegraphics[scale=.25]{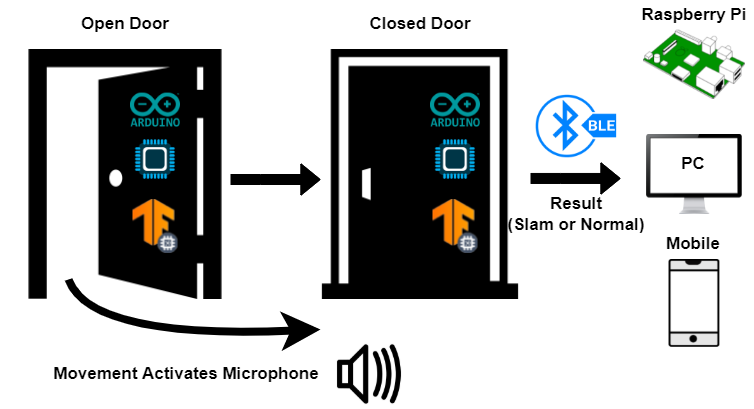}
	\caption{General overview of how the device could be deployed in the home, and how it can be used to detect if a door is being slammed or closed in an ordinary manner. The ML model is deployed to Arduino and runs inference.}
	\label{Fig:PARROTinterface}
\end{figure}

\begin{figure}
    \centering
	\includegraphics[scale=0.15 ]{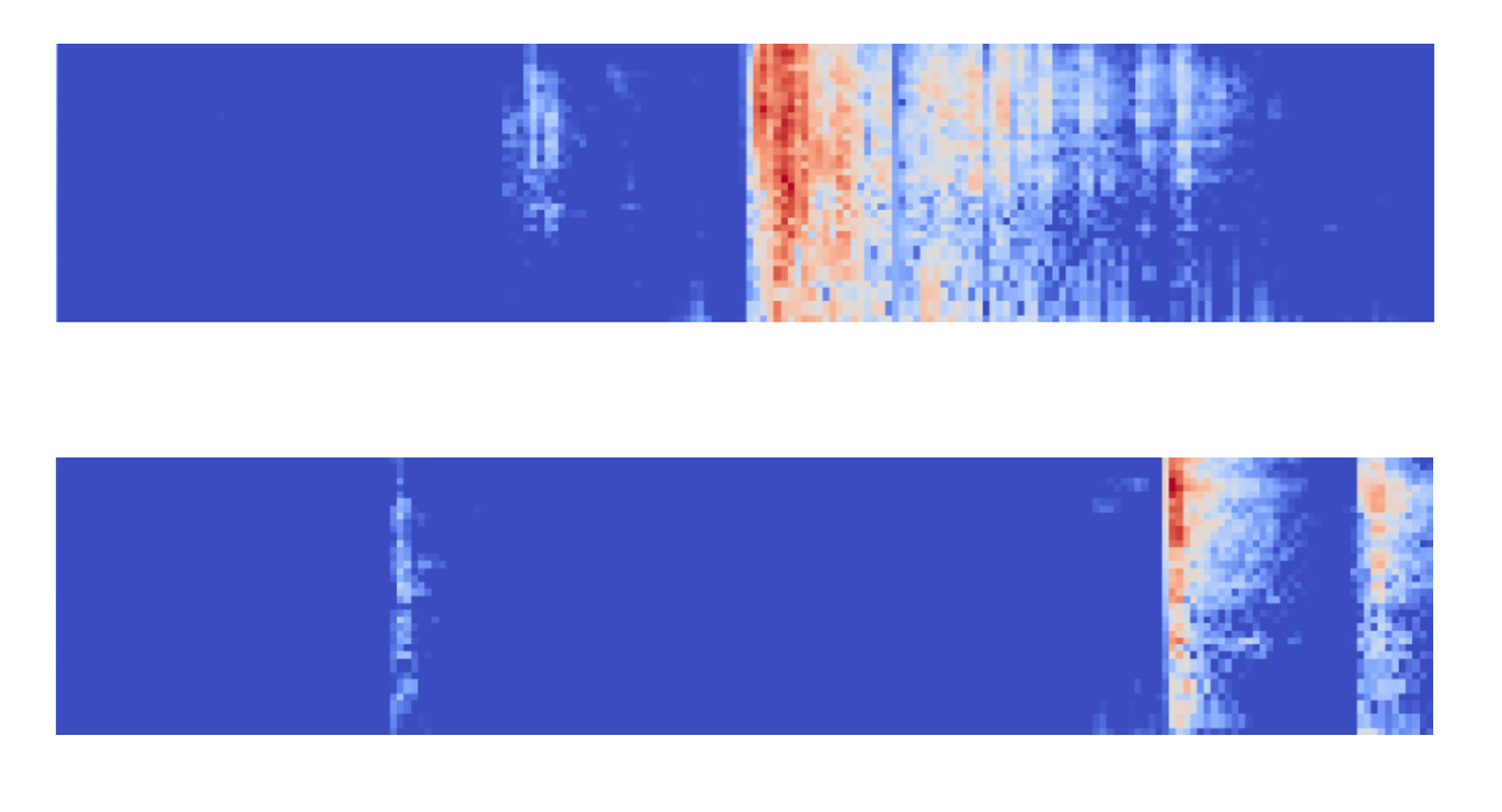}
	\caption{Example spectrograms produced via MFE on audio signal samples from the same door. A slam is shown above and a normal close below}
	\label{Fig:CGMwithPARROT}	
\end{figure}

\subsection{Deployment}
Once a model was developed and trained using the edge impulse toolkit, we needed to deploy the model in a sensible way on to a device. Edge Impulse provides a method for creating a custom Arduino library which includes the model and its supporting code for running inference, this can then be downloaded and saved to the appropriate file location for custom libraries. The code provided is general purpose and needs to be adapted further to suit the needs of a particular project. In our case, having the PDM microphone sample continuously would be undesirable. This would draw too much power and drain the batteries much faster, in addition to raising numerous privacy concerns linked to having an “always-on” mic in a private residence. As such we needed to devise some way of only activating the microphone, when necessary, via some trigger.

Given that any close of a door involves some degree of movement, accelerometer is an ideal choice for this task. By setting a threshold for absolute acceleration in terms of g, we can modify the code to only begin sampling once the threshold has been met or exceeded. Thereby avoiding the need for an always on microphone, in addition to the added benefit of filtering out non-violent slower closes of the door. Given that the Arduino would be fixed in a given door, one of the spatial axes, in our case Y, can be safely ignored as motion in this axis should not occur.

Once the trigger for the microphone is activated, the PDM mic on the board will commence sampling for a total of six seconds. By doing this we can eliminate unnecessary inference on irrelevant data, by extracting the relevant data from that six second period. Deleting non-relevant audio data gathered once movement of the door is detected should help reduce privacy concerns as well as helping avoid unnecessary classification of everyday door usage.

Once sampling and inference is complete, the predictions of the model can then be sent over BLE and read using free to download mobile applications such as nrfConnect (by Nordic Semiconductors). This eliminates the need for having a cable linking the board to a laptop, as was the case for  many of our earlier versions.

For effective use in the context of investigating domestic abuse and violence, it makes sense to deploy the device in a way that reduces the risk of tampering. As such the microcontroller can be embedded within the door itself, along with a power source that can be wirelessly recharged by authorised persons.


\end{document}